\documentclass{elsart}
\usepackage{doublespace}

\begin{document}

\raggedbottom

\begin{frontmatter}

\title{Compressing Probability Distributions}

\author{Travis Gagie}
\address{Department of Computer Science\\
	University of Toronto}

\begin{keyword}
Data compression
\end{keyword}

\end{frontmatter}

\section{Introduction}
\label{introduction}

Manipulating probability distributions is central to data compression, so
it is natural to ask how well we can compress probability distributions
themselves.  For example, this is useful for probabilistic
reasoning~\cite{BB03} and query optimization~\cite{JKMPSS98}. Our interest
in it stems from designing single-round asymmetric communication
protocols~\cite{AM01,Gag?}.  Suppose a server with high bandwidth wants to
help a client with low bandwidth send it a message; the server knows the
distribution from which the message is drawn but the client does not.  If
the distribution compresses well, then the server can just send that;
conversely, if the server can help the client in just one round of
communication and without sending too many bits, then the distribution
compresses well --- we can view the server's transmission as encoding it.

Compressing probability distributions must be lossy, in general, and it is
not always obvious how to measure fidelity.  In this paper we measure
fidelity using relative entropy because, in the asymmetric communication
example above, the relative entropy is roughly how many more bits we
expect the client to send with the server's help than if it knew the
distribution itself.  Let \(P = p_1, \ldots, p_n\) and \(Q = q_1, \ldots,
q_n\) be probability distributions over the same set.  Then the
\emph{relative entropy}~\cite{KL51} of $P$ with respect to $Q$ is defined
as
\[D (P \| Q) = \sum_{i = 1}^n p_i \log \frac{p_i}{q_i}\ .\]
By $\log$ we mean $\log_2$.  Despite sometimes being called
Kullback-Leibler distance, relative entropy is not a true distance metric:  
it is not symmetric and does not satisfy the triangle inequality.  
However, it is widely used in mathematics, physics and computer science as
a measure of how well $Q$ approximates $P$~\cite{CT91}.

We consider probability distributions simply as sequences of non-negative
numbers that sum to 1; that is, we do not consider how to store the sample
space.  In Section~\ref{algorithm_section} we show how, given a
probability distribution \(P = p_1, \ldots, p_n\), we can construct a
probability distribution \(Q = q_1, \ldots, q_n\) with \(D (P \| Q) < 2\)
and store $Q$ exactly in \(2 n - 2\) bits of space.  Constructing, storing
and recovering $Q$ each take \(O (n)\) time.  We also show how to trade
compression for fidelity and vice versa.  Finally, in
Section~\ref{data_structure_section}, we show how to store a compressed
probability distribution and query individual probabilities without
decompressing it.

\section{An Algorithm for Compressing Probability Distributions}
\label{algorithm_section}

The simplest way to compress a probability distribution $P$ is to
construct and store a Huffman tree~\cite{Huf52} for it.  This lets us
recover a probability distribution $Q$ with \(D (P \| Q) <
1\)~\cite{LG82,YY02} but takes both \(\Omega (n \log n)\) time and
\(\Omega (n \log n)\) bits of space. In this section, we show how to use
the following theorem, due to Mehlhorn~\cite{Meh77}, to compress $P$ by
representing it as a strict ordered binary tree.  A strict ordered binary
tree is one in which each node is either a leaf or has both a left child
and a right child.  We show how to trade compression for fidelity, by
applying this result repeatedly, or trade fidelity for compression, using
another approach.

\begin{thm}[Mehlhorn, 1977]
\label{mehlhorn_theorem}
Given a probability distribution \(P = p_1, \ldots, p_n\), we can
construct a strict ordered binary tree on $n$ leaves that, from left to
right, have depths less than \(\log (1 / p_1) + 2, \ldots, \log (1 / p_n)  
+ 2\).  This takes \(O (n)\) time.
\end{thm}

\begin{pf*}{PROOF SKETCH.}
For \(1 \leq i \leq n\), let
\[S_i = \frac{p_i}{2} + \sum_{j = 1}^{i - 1} p_j\ .\]
Consider the code in which the $i$th codeword is the first \(\lceil \log 
(2 / p_i) \rceil\) bits of the binary expansion of $S_i$; these bits 
suffice to distinguish $S_i$, so the code is prefix-free.  Notice the 
$i$th leaf of the corresponding code-tree has depth less than \(\log (1 / 
p_i) + 2\).
\qed
\end{pf*}

Once we have used Theorem~\ref{mehlhorn_theorem} to get a strict ordered
binary tree $T$, we store $T$.  It is important that $T$ be ordered;
otherwise, it would only store information about the multiset \(\{p_i\ :\
1 \leq i \leq n\}\), rather than the sequence $P$, so we would also need a
permutation on $n$ elements, which takes \(\Theta (n \log n)\) bits.

\begin{thm}
\label{algorithm_theorem}
Given a probability distribution \(P = p_1, \ldots, p_n\), we can 
construct a probability distribution \(Q = q_1, \ldots, q_n\) with 
\(\max_{1 \leq i \leq n} \{p_i / q_i\} < 4\), so \(D (P \| Q) < 2\), and 
store $Q$ exactly in \(2 n - 2\) bits of space.  Constructing, storing and 
recovering $Q$ each take \(O (n)\) time.
\end{thm}

\begin{pf}
We apply Theorem~\ref{mehlhorn_theorem} to $P$ to get a strict ordered
binary tree $T$ on $n$ leaves that, from left to right, have depths \(d_1,
\ldots, d_n\) with \(d_i < \log (1 / p_i) + 2\).  We store $T$ in \(2 n - 
2\) bits of space, represented as a sequence of balanced parentheses.

For \(1 \leq i \leq n\), let
\[q_i = \frac{2^{- d_i}}{\sum_j 2^{- d_j}}\ .\]
Since $T$ is strict, by the Kraft Inequality~\cite{Kra49},
\[\sum_{j = 1}^n 2^{- d_j} = 1\ ;\]
thus, \(q_i = 2^{- d_i} > p_i / 4\).
\qed
\end{pf}

Using Theorem~\ref{algorithm_theorem} as a starting point, we can improve
fidelity at the cost of using more space.  One approach is given below; we 
leave as future work finding better tradeoffs.

\begin{thm}
\label{tradeoff_theorem}
Given a probability distribution \(P = p_1, \ldots, p_n\) and an integer
\(k \geq 2\), we can construct a probability distribution \(Q = q_1,
\ldots, q_n\) with \(\max_{1 \leq i \leq n} \{p_i / q_i\} < 2 +
\frac{1}{2^{k - 3}}\), so \(D (P \| Q) < \log \left( 2 + \frac{1}{2^{k -
3}} \right)\), and store $Q$ exactly in \(k n - 2\) bits of space.  
Constructing, storing and recovering $Q$ each take \(O (k n)\) time.
\end{thm}

\begin{pf}
By induction on $k$.  By Theorem~\ref{algorithm_theorem}, the claim is 
true for \(k = 2\).  Let \(k \geq 3\) and assume the claim is true for \(k 
- 1\).

Let \(Q' = q_1', \ldots, q_n'\) be the probability distribution we
construct when given $P$ and \(k - 1\).  Let \(B = b_1 \cdots b_n\) be the 
binary string with \(b_i = 1\) if \(p_i / q_i' \geq 1 + \frac{1}{2^{k - 
3}}\) and \(b = 0\) otherwise.  For \(1 \leq i \leq n\), let
\[q_i = \left\{ \begin{array}{ll}
	\frac{2 q_i'}{\sum_{b_i = 1} 2 q_i' + \sum_{b_i = 0} q_i'}
		\hspace{0.25in} \mbox{}
		& \mbox{if \(b_i = 1\), and} \\
	\frac{q_i'}{\sum_{b_i = 1} 2 q_i' + \sum_{b_i = 0} q_i'}
		& \mbox{if \(b_i = 0\).}
\end{array} \right.\]
Notice we can store $Q$ exactly in \(k n - 2\) bits of space, using \((k 
- 1) n - 2\) bits for $Q'$ and $n$ bits for $B$.  Also,
\[\sum_{b_i = 1} 2 q_i' + \sum_{b_i = 0} q_i'
= \sum_{i = 1}^n q_i' + \sum_{b_i = 1} q_i'
\leq 1 + \sum_{b_i = 1} \frac{p_i}{1 + 1 / 2^{k - 3}}
\leq \frac{2^{k - 2} + 1}{2^{k - 3} + 1}\ .\]

If \(b_i = 1\), then \(q_i \geq 2 q_i' \cdot \frac{2^{k - 3} + 1}{2^{k - 
2} + 1}\).  Since, by assumption, \(q_i' > \frac{p_i}{2 + 1 / 2^{k - 
4}}\), we have
\[q_i
> \frac{2 p_i}{2 + 1 / 2^{k - 4}} \cdot \frac{2^{k - 3} + 1}{2^{k - 2} + 1}\]
and so \(p_i / q_i < 2 + \frac{1}{2^{k - 3}}\).  If \(b_i = 0\), then
\(q_i' > \frac{p_i}{1 + 1 / 2^{k - 3}}\).  Since \(q_i \geq q_i' \cdot
\frac{2^{k - 3} + 1}{2^{k - 2} + 1}\), we have
\[q_i
> \frac{p_i}{1 + 1 / 2^{k - 3}} \cdot \frac{2^{k - 3} + 1}{2^{k - 2} + 1}\]
and so \(p_i / q_i < 2 + \frac{1}{2^{k - 3}}\).

By assumption, constructing $Q'$ takes \(O ((k - 1) n)\) time and 
constructing $B$ and $Q$ from $Q'$ takes \(O (n)\) time.  Thus, 
constructing, storing and recovering $Q$ each take \(O (k n)\) time.
\qed
\end{pf}

It may be possible to strengthen Theorem~\ref{algorithm_theorem} using
results about alphabetic Huffman codes (e.g.,~\cite{She92}).  We base it
on Theorem~\ref{mehlhorn_theorem} for two reasons: Mehlhorn's construction
takes \(O (n)\) time, whereas known algorithms for constructing alphabetic
Huffman codes take \(\Omega (n \log n)\) time~\cite{KM95}, and the
guarantee that \(\max_{1 \leq i \leq n} \{p_i / q_i\} < 4\) makes the 
proof of Theorem~\ref{tradeoff_theorem} cleaner.

Using a different approach, we can also reduce the space used at the cost
of reducing fidelity.

\begin{thm}
\label{ratio_theorem}
Given a probability distribution \(P = p_1, \ldots, p_n\) and \(c \geq
1\), we can construct a probability distribution \(Q = q_1, \ldots, q_n\)
with \(D (P \| Q) \leq c \cdot H (P) + \log (\pi^2 / 3)\) and store $Q$
exactly in at most \(\lfloor n^{1 / (c + 1)} \rfloor (\lfloor \log n 
\rfloor + 1)\) bits of space.  Constructing, storing and recovering $Q$ 
each take \(O (n)\) time.
\end{thm}

\begin{pf}
Let \(t \leq \lfloor n^{1 / (c + 1)} \rfloor\) be the number of 
probabilities in $P$ that are at least \(\frac{1}{n^{1 / (c + 1)}}\).  Let 
\(r_1, \ldots, r_t\) be such that \(p_{r_j}\) is the $j$th largest 
probability in $P$, and let \(R = \{r_1, \ldots, r_t\}\).  Thus,
\[p_{r_1} \geq \cdots \geq p_{r_t}
\geq \frac{1}{n^{1 / (c + 1)}}
> \max_{i \not \in R} \{p_i\}\ .\]
Computing the set $R$ takes \(O (n)\) time and sorting it takes \(O (n^{1
/ (c + 1)} \log n) \subset O (n)\) time.  For \(1 \leq j \leq t\), let 
\(q_{r_j} = 3 / (\pi j)^2\); since
\[\sum_{j = 1}^{\infty} \frac{1}{j^2} = \frac{\pi^2}{6}\ ,\]
we have
\[\sum_{j = 1}^t q_{r_j} < \frac{1}{2}\ .\]
For \(i \not \in R\), let
\[q_i
= \frac{1 - \sum_{j = 1}^t 3 / (\pi j)^2}{n - t}
> \frac{1}{2 n}\ .\]
Storing $Q$ as the binary representations of \(r_1, \ldots, r_t\) takes at
most \(\lfloor n^{1 / (c + 1)} \rfloor\) \((\lfloor \log n \rfloor + 1)\)  
bits of space and \(O (n)\) time.

For \(1 \leq j \leq t\), since $p_{r_j}$ is the $j$th largest probability 
in $P$, we have \(p_{r_j} \leq 1 / j\).  Therefore,
\begin{eqnarray*}
H (P)
& = & \sum_{i = 1}^n p_i \log \frac{1}{p_i} \\
& \geq & \sum_{j = 1}^t p_{r_j} \log j +
	\sum_{i \not \in R} p_i \log n^{1 / (c + 1)} \\
& = & \sum_{j = 1}^t p_{r_j} \log j +
	\frac{\sum_{i \not \in R} p_i \log n}{c + 1}\ .
\end{eqnarray*}
Compare this with
\begin{eqnarray*}
D (P \| Q)
& = & \sum_{i = 1}^n p_i \log \frac{p_i}{q_i} \\
& = & \sum_{j = 1}^t p_{r_j} \log \frac{1}{q_{r_j}} +
	\sum_{i \not \in R} p_i \log \frac{1}{q_i} - H (P) \\
& \leq & \sum_{j = 1}^t p_{r_j} \log \frac{(\pi j)^2}{3} +
	\sum_{i \not \in R} p_i \log (2 n) - H (P) \\
& = & 2 \sum_{j = 1}^t p_{r_j} \log j +
	\sum_{i \not \in R} p_i \log n +
	\log \left( \frac{\pi^2}{3} \right) \sum_{j = 1}^t p_{r_j} +
	\sum_{i \not \in R} p_i - H (P) \\
& \leq & 2 \sum_{j = 1}^t p_{r_j} \log j +
	\sum_{i \not \in R} p_i \log n +
	\log \frac{\pi^2}{3} - H (P)\ .
\end{eqnarray*}
Since \(c \geq 1\),
\[\frac{D (P \|Q) - \log (\pi^2 / 3)}{H (P)} + 1
\leq \frac{2 \sum_{j = 1}^t p_{r_j} \log j +
	\sum_{i \not \in R} p_i \log n}
	{\sum_{j = 1}^t p_{r_j} \log j +
	\frac{\sum_{i \not \in R} p_i \log n}{c + 1}}
\leq c + 1\ ;\]
that is, \(D (P \| Q) \leq c \cdot H (P) + \log (\pi^2 / 3)\).
\qed
\end{pf}

If space is at a premium, we may need to work with $Q$ without
decompressing it.  Notice we can do this by storing \(\{(r_1, 1), \ldots,
(r_t, t)\}\) in order by first component, which takes at most
\[\lfloor n^{1 / (c + 1)} \rfloor \left( \lfloor \log n \rfloor + 
	\left\lfloor \frac{\log n}{c + 1} \right\rfloor + 2 \right)\]
bits of space.  Given $i$ between 1 and $n$, we can compute
\[q_i = \left\{ \begin{array}{ll}
3 / (\pi j)^2
	& \hspace{0.25in} \mbox{if \(i = r_j\),} \\
\frac{1 - \sum_{j = 1}^t 3 / (\pi j)^2}{n - t}
	& \hspace{0.25in} \mbox{if \(i \not \in \{r_1, \ldots, r_t\}\)}
\end{array} \right.\]
in \(O (\log t) \subseteq O (\log (n) / c)\) time.

\section{A Data Structure for Compressed Probability Distributions}
\label{data_structure_section}

In this section, we show how to work with a probability distribution
compressed with Theorem~\ref{algorithm_theorem} without decompressing it,
using a succinct data structure due to Munro and Raman~\cite{MR01}.  This
data structure stores a strict ordered binary tree on $n$ leaves in \(2 n
+ o (n)\) bits of space and supports queries that, given a node, return
its parent, left child, right child and number of descendants.  Each of
these queries takes \(O (1)\) time.  Notice that, given $i$ between 1 and
$n$, we can find the depth $d$ of the $i$th leaf in \(O (d)\) time.

\begin{thm}
\label{data_structure_theorem_1}
Given a probability distribution \(P = p_1, \ldots, p_n\), we can
construct a data structure that uses \(2 n + o (n)\) bits of space and
supports a query that, given $i$ between 1 and $n$, returns $q_i$ in \(O
(\log (1 / q_i))\) time.  Here, \(Q = q_1, \ldots, q_n\) is a probability
distribution with \(\max_{1 \leq i \leq n} \{p_i / q_i\} < 4\), so \(D (P
\| Q) < 2\) and \(O (\log (1 / q_i)) \subseteq O (\log (1 / p_i))\).
\end{thm}

\begin{pf}
As for Theorem~\ref{algorithm_theorem}, but with the sequence of balanced 
parentheses replaced by an instance of Munro and Raman's data structure.
\qed
\end{pf}

A drawback to Theorem~\ref{data_structure_theorem_1} is that querying a
very small probability might take \(\Theta (n)\) time.  We can fix this by
smoothing the given probability distribution slightly.

\begin{thm}
\label{data_structure_theorem_2}
Given a probability distribution \(P = p_1, \ldots, p_n\) and \(\epsilon 
> 0\), we can construct a data structure that uses \(2 n + o (n)\) bits of 
space and supports a query that, given $i$ between 1 and $n$, returns 
$q_i$ in \(O (\log (1 / q_i))\) time.  Here, \(Q = q_1, \ldots, q_n\) is a 
probability distribution with \(D (P \| Q) < 2 + \epsilon\) and \(\log (1 
/ q_i) \in O (\log \min (1 / p_i, n / \epsilon))\).
\end{thm}

\begin{pf}
Let \(P' = p_1', \ldots, p_n'\), where
\[p_i'
= \frac{p_i}{1 + \epsilon / 4} + \frac{\epsilon / 4}{(1 + \epsilon / 4) n}\ .\]
We apply Theorem~\ref{data_structure_theorem_1} to $P'$; let $Q$ be the 
stored distribution.  Notice
\[\max_{1 \leq i \leq n} \left\{ \frac{p_i}{q_i} \right\}
\leq \max_{1 \leq i \leq n} \left\{ \frac{p_i}{p_i'} \right\} \cdot
	\max_{1 \leq i \leq n} \left\{ \frac{p_i'}{q_i} \right\}
\leq \left( 1 + \frac{\epsilon}{4} \right) \cdot 4
= 4 + \epsilon\ .\]
Since $\log$ is convex, it follows that \(D (P \| Q) < 2 + \epsilon\).  
Since
\[q_i
> \max \left( \frac{p_i}{4 + \epsilon},
	\frac{\epsilon}{4 n} \right)\ ,\]
we have \(\log (1 / q_i) \in O (\log \min (1 / p_i, n / \epsilon))\).
\qed
\end{pf}

\bibliographystyle{plain}
\bibliography{distributions}

\end{document}